\documentclass[prl,twocolumn,showpacs]{revtex4}

\usepackage{graphicx}
\usepackage{dcolumn}
\usepackage{bm}

\begin{document}

\title{Development of correlated quasiparticle conductance peak as molecule-linked gold nanoparticle films transition from Mott-insulator to metal phases }

\author{Patrick Joanis}
\author{Monique Tie}
\author{Al-Amin Dhirani}
\email{adhirani@chem.utoronto.ca} \affiliation{Department of
Chemistry, University of Toronto, Toronto, Ontario, Canada, M5S 3H6}

\date{\today}

\begin{abstract}
We have studied conductance ($\it{g}$) of butanedithiol-linked gold
nanoparticle films across a percolation insulator-to-metal
transition. As the transition proceeds, electrons become itinerant
(i.e. Coulomb charging and kinetic effects are both significant),
and films exhibit a previously unobserved zero-bias conductance peak
(ZBCP). The peak is much more pronounced and easily observed using
electromigration-induced break junction (BJ) contacts rather than
macroscopic 4-probe electrodes. We attribute this ZBCP to quantum
correlations amongst electrons, in view of other temperature-
($\it{T}$-) and magnetic ($\it{B}$-) dependent measurements as well
as predictions of the Hubbard model and dynamic mean field theory in
this transition regime. Metallic film resistances ($\it{R}$'s)
increase linearly with $\it{T}$, but with suggested scattering
lengths that, anomalously, are shorter than inter-atomic distances.
Similar so-called ``bad-metallic'' behaviour has been observed in
several studies of correlated systems, and is still being
understood. We find here that the anomalous $\it{R}$ behaviours are
associated with the ZBCP. This system can serve as a new test bed
for studying correlated electrons and points to a nano
building-block strategy for fashioning novel correlated materials.
\end{abstract}


\maketitle

Transitions where charges become itinerant while Coulomb
interactions remain strong appear in a number of exotic materials,
such as vanadium oxides that exhibit huge resistivity changes across
an insulator-to-metal transition, cuprates that exhibit high Tc
superconductivity and manganites that exhibit colossal
magnetoresistance \cite{DMFT}. They have been intensely studied, and
are known to arise in materials with partially open $\it{d}$ and
$\it{f}$ orbitals, whose confined nature leads to strong electronic
repulsion.  As such materials transition from Mott insulating to
metallic states, electrons interact with each other strongly and,
therefore, move in a correlated fashion. Calculations combining a
local density approximation with dynamic mean field theory
\cite{DMFT} predict that as the transition proceeds, the upper and
lower Hubbard bands become more closely spaced, and a quasiparticle
peak develops at the Fermi level. Experimentally, a quasiparticle
peak has been observed in vanadium oxide via optical \cite{PUDDLES2}
and photoemission \cite{photoemission peak} spectroscopy and in
NiS$_{2-x}$Se$_{x}$ via tunneling spectroscopy \cite{tunneling
peak}.

Films of Au nanoparticles (NPs) cross-linked with $n$-alkanedithiol
(HS(CH$_2$)$_n$SH, e.g. taking $n$ = 4) can exhibit a
thickness-driven insulator-to-metal percolation transition in which
both Coulomb and kinetic effects can be significant, suggesting that
this system can potentially serve as a novel and potentially
advantageous test bed for studying electron correlations \cite{NPF,
SYNTHESIS, MORE_NPF}. Use of prefabricated NPs with relatively
narrow size distribution limits averaging \cite{HEATH}. Further,
except for a small fraction of NPs that self-assemble to the
functionalized surface, nanoparticle films (NPFs) grow by NPs
attaching to one another, forming metallic pathways in a
fractal-like rather than 2-D manner. As a result, the percolation
transition is gradual, and phases that are intermediate between
insulating and metallic extremes are more easily observed.

In the present study we explore the evolution of conductance
($\it{g}$) for butanedithiol-linked NPFs as a function of
temperature ($\it{T}$) and bias ($\it{V}$) across the percolation
insulator-to-metal transition. Previous studies of other systems
have shown that as the transition proceeds, first, nanoscale
metallic puddles with exotic quasiparticles form \cite{PUDDLES1,
PUDDLES2}. In order to better observe these exotic phases, we use
break junctions (BJs) to interrogate nanoscale electrical properties
of the NPFs.  We also studied macroscopic properties of films using
four-probe electrodes. As the NPFs become metallic, both BJ and
four-probe samples exhibit a zero-bias conductance peak (ZBCP) that
we attribute to correlated quasiparticles. We find that scattering
lengths, estimated assuming Boltzman transport, are smaller than
inter-atomic separations, implying ``bad metallic'' behaviour that
has been attributed to correlation effects in other systems but
whose origin has been unresolved. As with other bad metals, NPF
$\it{R}$'s increases linearly with $\it{T}$. We find here that this
behaviour can be traced to the ZBCP.

Both four-probe and BJ samples were fabricated on glass substrates
 that were first cleaned by
immersion in hot piranha for 30 min and then functionalized by
immersion in a boiling toluene solution of
3-mercaptopropyltrimethoxysilane for 20 min. For BJ samples, a Au
wire that was a 100 ${\mu}$m wide, 8 mm long, 16 nm thick in the
middle (over a 100 ${\mu}$m stretch), and 150 nm thick over the rest
of its length was deposited using metal evaporation and shadow
masks. A ramped voltage was applied in L-N$_2$ until the wire broke
due to electromigration. BJs exhibited gaps with widths that ranged
from a few $\mu$m to a few nm (estimated from measurements of tunnel
currents). For four-probe samples, Au electrodes $\sim$ 6 mm long,
200 nm thick and separated by 2 mm were employed. For both types of
samples, copper magnet wires were soldered to the Au electrodes
using indium before film self-assembly. Au NPs 5.0$\pm$ 0.8 nm in
diameter were synthesized using published methods \cite {SYNTHESIS},
and butanedithiol used to link Au NPs was purchased and used as
received.   NPFs were self-assembled by alternately immersing the
slides in an Au NP solution for 10-60 min and a 0.5 mM ethanolic
butanedithiol solution 10 min until a desired $\it{R}$ was reached.
$\it{g}$ at various $\it{V}$'s were determined using lockins and at
various $\it{T}$'s using a Quantum Design PPMS. To confirm
reproducibility, the evolution of four BJ and a pair of four-probe
samples were studied in detail, each at several NPF thicknesses.

\begin{figure}
\includegraphics{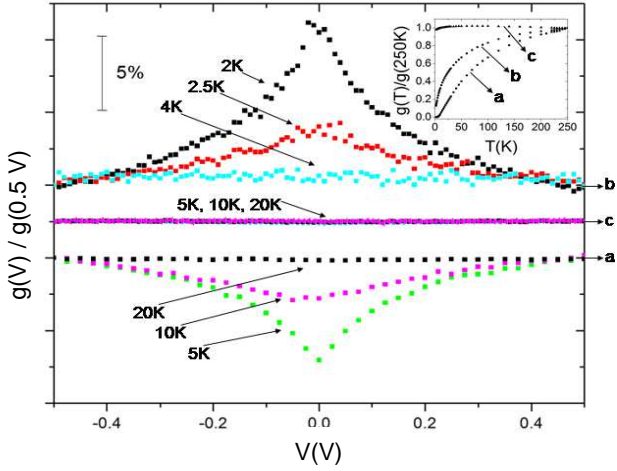}
\caption{\label{1} $\it{g}$ vs. $\it{V}$ at different $\it{T}$ (main
panel) and $\it{g}$ vs. $\it{T}$ at $\it{V}$ = 0 (inset) obtained
using a four-probe electrode configuration. All the data were
obtained using the same sample at different film thicknesses. Data
in the main panel and inset with the same labels correspond to the
same sample thickness, namely 12-, 18- and 25- one-hour exposures to
Au NP solution, labeled ``a'' to ``c'', respectively, in order of
increasing thickness.  The data in the main panel for a given
thickness and $\it{T}$ have been normalized to their respective
values at 0.5 V and offset for clarity. }
\end{figure}

Figure 1 plots $\it{g}$ vs. $\it{T}$ and $\it{V}$ for a four-probe
sample. The data were obtained using the same sample at three
different NP/dithiol exposure cycles corresponding to behaviour that
is insulating, just metallic and well metallic - graphs labeled
``a'', ``b'' and ``c'', respectively. In the insulating regime,
$\it{g}$ $\rightarrow$ 0 as $\it{T}$ $\rightarrow$ 0 and,
concurrently, $\it{g}$ is suppressed near zero volts. This
suppression is observed in both two- and four-probe measurements
performed simultaneously on the same sample, indicating that at this
thickness, film $\it{R}$ dominates over contact $\it{R}$.  When
thickness increases so the sample is just at the transition,
remarkably the four-probe data exhibit a ZBCP at low $\it{T}$. The
two-probe data continue to show a small suppression of $\it{g}$ as
before (data not shown). These data indicate that the ZBCP observed
in the four-probe data is a film effect and that the suppression
observed in the two-probe data arises from a small barrier at the
contact(s). As $\it{T}$ increases (beyond 2K to 10K depending on the
sample), the ZBCP wanes and $\it{g}$ becomes independent of $\it{V}$
(ohmic). As film thickness increases still further, four-probe
$\it{g}$ data behave essentially ohmically even at the lowest
accessible $\it{T}$. Simultaneously measured two-probe $\it{g}$ data
continue to exhibit the presence of a small barrier.

Insulating film behaviour is consistent with single electron
charging phenomena (Coulomb blockade) \cite {CB}. Below a
percolation threshold, electrons tunnel on to and must charge NPs or
metallic clusters of NPs. Previous studies have shown that
single-electron charging energies increase with decreasing particle
size and can become substantial at nm length scales \cite {CB}; e.g.
the single electron charging energy of a 1 nm sphere can be
estimated using Coulomb's law as $\sim$1 eV.  This Coulomb barrier
leads to suppression of $\it{g}$ at low $\it{V}$ and $\it{T}$ in
NPFs.

A variety of processes are known to cause ZBCPs. Reflectionless
tunneling can be excluded here because it is a contact rather than a
film phenomenon, while ZBCP appear in measurements of film $\it{g}$.
 Lack of Zeeman splitting excludes the Kondo effect.  The field ($\it{B}$) at which
Zeeman peak splitting would be expected for a zero-bias anomaly
(Kondo effect) can be estimated using the Haldane relation:
$\it{2\mu_{B}B} \sim
 \it{k_{B}T_{K}} = \sqrt{ \Gamma \it{U}}/2 \cdot exp [ -\pi\epsilon_{0}
(-\epsilon_{0}+\it{U} )/\Gamma \it{U}] $ where $\mu_{B}$ is the Bohr
magneton, $\epsilon_{0}$ is the energy level through with transport
occurs, $\it{U}$ is the on-site repulsion, and $\it{\Gamma}$ is the
energy broadening \cite{KONDO, Haldane}. Since there are no magnetic
impurities in our system (otherwise a Kondo peak would be observed
even in thick films), we attribute any potential Kondo effect to a
mechanism based on single electron charging.  An Arrhenius plot of
$\it{g}$ vs. $\it{T}$ for a non-metallic film (Fig. 2a, inset)
yields a temperature scale of 30 K. Using this as an estimate for
$\epsilon_{0}$, $\it{U} \sim 2\epsilon_{0}$, and $\Gamma \sim
\epsilon_{0}$, splitting should then be observed for $\it{B}\sim$
3T. We applied $\sim$3-5 times this field, but did not observe peak
splitting, even with $<$1\% sensitivity to changes in $\it{g}$.
Joule heating can also be excluded because $\it{g}$ increases with
$\it{T}$ (Fig. 1, inset). Heating caused by increasing $\it{V}$
should cause $\it{g}$ to increase and should generate a zero-bias
conductance suppression, rather than a peak as observed.

\begin{figure}
\includegraphics{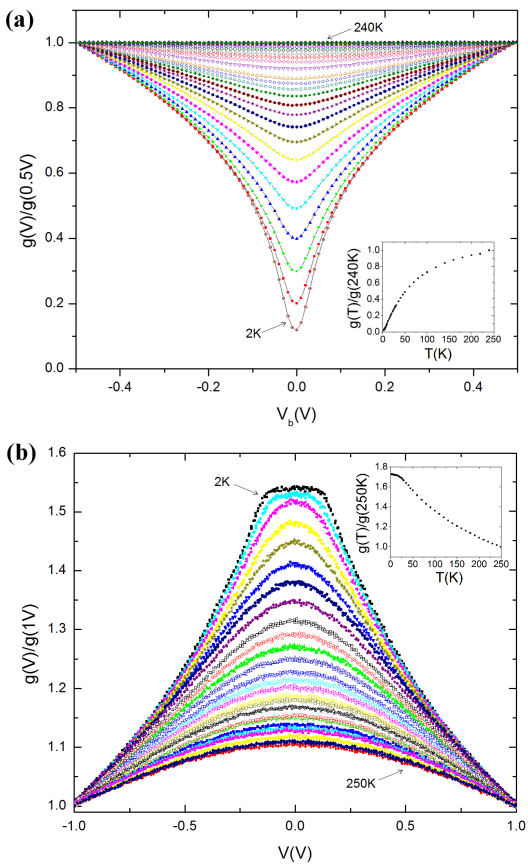}
\caption{\label{2} Evolution of $\it{g}$ measured using a BJ as a
NPF crosses the insulator-to-metal transition. a. $\it{g}$ vs.
$\it{V}$ at different $\it{T}$'s from 2 K to 240 K (main panel) and
 $\it{g}$ vs. $\it{T}$ at $\it{V}$ = 0 (inset) for an insulating, 10
dithiol/NPF. Data for a given $\it{T}$ have been normalized to their
respective values at 0.5 V.  The conductance at 250 K is 9.39 x
10$^{-5} \Omega^{-1}$ .  Data were obtained at 2K to 30K in 1K
steps, 30K to 50K in 5K steps, 50K to 100K in 10K steps and 100K to
240K in 20K steps.  b. $\it{g}$ vs. $\it{V}$ at different $\it{T}$'s
from 2 K to 250 K (main panel) and $\it{g}$ vs. $\it{T}$ at $\it{V}$
= 0 (inset) for the same sample as in a. with a thicker (50
dithiol/nanoparticle exposure) film that has turned metallic. Data
for a given temperature have been normalized to their respective
values at 0.5 V.  The conductance at 250 K is 0.02 $\Omega^{-1}$ .
Data were obtained at 2K to 30K in 1K steps, 30K to 50K in 5K steps,
and 50K to 250K in 10K steps.}
\end{figure}

Using BJs to measure $\it{g}$ improves resolution of spectroscopic
features and reveals a mesoscale origin of the ZBCP. Figure 2 shows
nanoscale film behaviour as the insulator-to-metal transition
progresses. $\it{g}$ vs. $\it{V}$ data for an insulating NPF (Fig.
2a, main panel) show strong Coulomb blockade suppression near 0 V at
2 K, and zero-bias $\it{g}$ vs. $\it{T}$ data (inset) show that
$\it{g}$ $\rightarrow$ 0 as $\it{T}$ $\rightarrow$ 0. As the film
was subjected to more NP/dithiol immersion cycles, it eventually
became metallic ($\it{g}$ $\rightarrow$ non-zero constant as
$\it{T}$ $\rightarrow$ 0) but with a persistent small energy barrier
since $\it{g}$ was observed to increase with both $\it{V}$ and
$\it{T}$ (data not shown). When $\it{V}$ was increased to a few
volts at low $\it{T}$, $\it{g}$ suddenly and irreversibly increased
10- to 100-fold remaining stable thereafter, and the (contact)
barrier was no longer observed.   All BJ samples with sufficiently
thick NPF bridges exhibited this break down after which pronounced
ZBCPs became apparent.  The peaks exhibit an Ohmic plateau region at
low $\it{V}$ (Fig. 2b, main panel) and survive to high $\it{T}$'s -
at least 250 K, beyond which the data tend to become noisy.  The
significantly larger energy scale provided by the voltage ($\sim$
eV) arises from peak broadening due to voltage division across the
film.  As more dithiol/NP is self-assembled, eventually $\it{g}$'s
and peak widths change only slightly, suggesting that most of the
current flows through a portion of the NPF located in the gap
between Au electrodes.   We attribute the ZBCPs to the formation of
a ``meso-metallic'' phase since 1) they are associated with non-zero
$\it{g}$ at low $\it{T}$ (i.e. films are metallic - see Fig. 2b,
inset); 2) $\it{g}$ is not entirely ohmic (the peaks have an ohmic
plateau below a threshold $\it{V}$ but decreasing $\it{g}$ above);
and 3) the ZBCPs are more pronounced in BJ samples than in
macroscopic four-probe samples. This attribution is also consistent
with dynamic mean field calculations using the Hubbard model that
predict the development of a peak in the density of states at the
Fermi level at the onset of metallicity.

\begin{figure}
\includegraphics{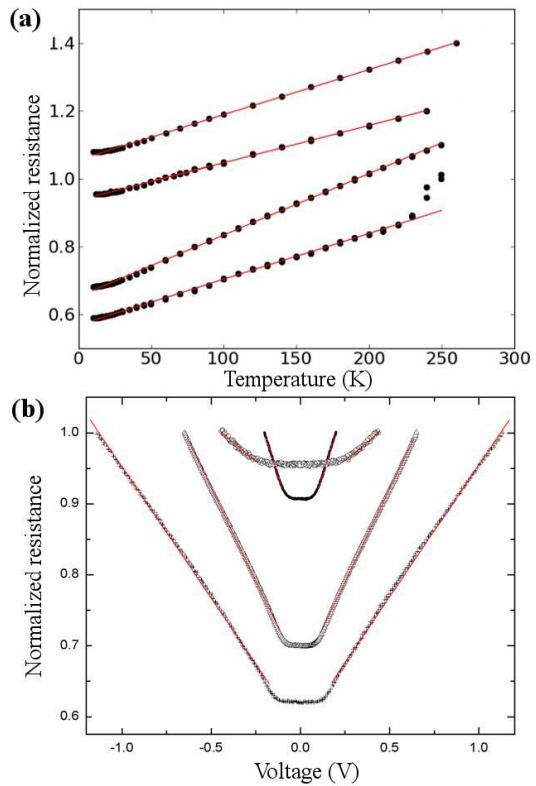}
\caption{\label{3} $\it{R}$ vs. $\it{T}$ and vs. $\it{V}$ data for
metallic NPFs measured using BJs. a. $\it{R}$ vs. $\it{T}$ at
$\it{V}$ = 0 for four metallic NPFs. b. $\it{R}$ vs. $\it{V}$ at 4 K
for the same four samples as in a. Data are fit to straight lines
over a portion of their respective ranges as shown. All data have
been normalized using their respective maximum values and offset in
a. for clarity.}
\end{figure}

Metallic states of various correlated materials - such as organic
conductors, alkali-doped C60 as well as those of various oxides
(including cuprates, manganites and vanadium dioxide) - exhibit a
linear increase of $\it{R}$ with $\it{T}$ and quasiparticle
scattering lengths that, remarkably, are smaller than interatomic
distances. Similar behaviours are observed here (see below).  Such
``bad metallic'' behaviours are still being elucidated in the
literature \cite{PUDDLES2, BADMETAL}. It is generally agreed,
however, that the quasiparticle picture of a particle with a well
defined momentum on the order of the Fermi momentum experiencing
occasional scattering breaks down in these systems as a result of
strong electron-electron correlations. Figures 3a and 3b plot
zero-bias $\it{R}$ vs. $\it{T}$ and $\it{R}$ vs. $\it{V}$ data for
four BJ samples. The data indicate that $\it{T}$ independent
scattering dominates $\it{R}$ in these NPFs. Given the nanogranular
nature of NPFs, ubiquitous $\it{T}$ independent elastic scattering
from NP interfaces might be expected. However, assuming uncorrelated
electrons that scatter occasionally (Boltzman transport) for
arguments sake, the resistivity would be given by $\rho =
3\pi^2\hbar/e^2k_{F}^2l$, where $\rho$ is the resistivity, $\hbar$
is Planck's constant, $e$ is the quantum of charge, $k_F$ is the
Fermi wave vector and $l$ is the scattering length \cite{mean free
path}. For our BJ samples, taking a typical $\it{R}$ of $\sim$ 50
$\Omega$ , a sample area of 150 $\mu$m x 100 nm, a sample length of
100 nm, and assuming a Fermi wavelength of 0.54 nm for gold, we
estimate $l \sim 10^{-3}$ ${\AA}$. For our four-probe samples,
taking a typical $\it{R}$ of $\sim$ 200 $\Omega$, a sample area of 3
mm x 100 nm, a sample length of 3 mm, and assuming a Fermi
wavelength of 0.54 nm, we estimate $l \sim 0.3 {\AA}$. These are
rough estimates because of uncertainties in film thickness, but it
is safe to conclude that the predicted scattering lengths in the
nanoscale BJ samples are smaller than interatomic distances and
smaller than those for the macroscopic samples.   This increased
tendency for bad metallic behaviour at smaller length scales is
correlated with the ZBCP being more pronounced in the BJ samples.
This is also consistent with scan probe studies by others that have
shown that correlated materials are heterogeneous \cite{PUDDLES1,
PUDDLES2, HETERO}. As the insulator-to-metal transition proceeds,
nanoscale metallic puddles form and grow; these regions are not
smaller versions of fully formed bulk metallic state.

\begin{figure}
\includegraphics{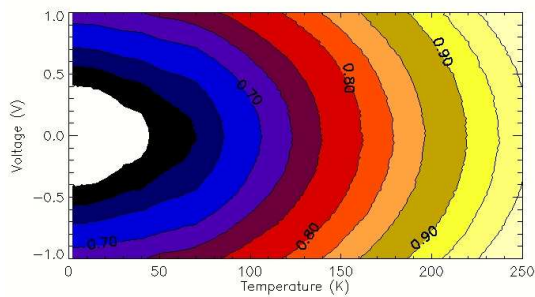}
\caption{\label{4} A contour plot of $\it{R}$ at various $\it{V}$s
and $\it{T}$s using the same data shown in Fig. 3. The data have
been normalized to their maximum $\it{R}$, and contour labels denote
fractions of this maximum. }
\end{figure}

Correlated systems, e.g. the cuprates, in addition to exhibiting
anomalously large resistance, exhibit $\it{R}$ that increases
linearly with $\it{T}$ to very high $\it{T}$. This behaviour is not
well understood \cite{BADMETAL}.  Figures 3a and b show that both
$\it{R}$ vs. $\it{T}$ at zero-bias and $\it{R}$ vs. $\it{V}$ at low
$\it{T}$ are constant initially and then both vary linearly for all
four BJ samples. Figure 4 plots contour lines for $\it{R}$ vs.
$\it{V}$ and $\it{T}$ data for one of the samples. Equally spaced
contour lines reflect the underlying linearity of $\it{R}$ as a
function of either variable.  That is, this study suggests that both
behaviours can be traced back to the quasiparticle peak in the
density of states. Further theoretical work modeling transport
measurements and exploring this link would be beneficial. Another
significant implication of the present study is that it suggests
that nanostructured materials constitute a novel class of correlated
materials since Coulomb interactions in nanostructures are
inherently large and, as electrons become itinerant, both Coulomb
and kinetic effects should be significant. Such materials can
provide a new, controllable and potentially rich platform to study
these exotic phenomena since there is a tremendously wide and,
indeed, designable selection of nanostructures that can be used as
material building blocks.

We thank Y.-B. Kim for discussions.  This work was supported by the
Natural Science and Engineering Research Council for Canada.


\begin{thebibliography}{30}

\bibitem{DMFT}
G. Kotliar and D. Vollhardt, Physics Today March, 53 (2004). G.
Kotliar et al., Rev. Mod. Phys. \textbf{78}, 865 (2006). K. Held,
Adv. Phys. \textbf{56}, 829 (2007). M. Jarrell, Phys. Rev. Lett.
\textbf{69}, 168 (1992).


\bibitem{PUDDLES2}
M. M. Qazilbash et al., Science \textbf{318}, 1750 (2007).


\bibitem{photoemission peak}
S.-K. Mo et al., Phys. Rev. Lett. \textbf{90}, 186403 (2003).

\bibitem{tunneling peak}
K. Iwaya et al., Phys. Rev. B \textbf{70}, 161103R (2004).

\bibitem{NPF}
M. Brust et. al, Adv. Mater. \textbf{7}, 795 (1995). A.
Zabet-Khosousi, P. E. Trudeau, Y. Suganuma, A.-A. Dhirani and B.
Statt, Phys. Rev. Lett. \textbf{96}, 156403 (2006).

\bibitem{SYNTHESIS}
N. Fishelson et al, Langmuir \textbf{17}, 403 (2001).

\bibitem{MORE_NPF}
M. Brust and C. J. Kiely, Colloids Surf. A \textbf{202}, 175 (2002).
A. Zabet-Khosousi and A.-A. Dhirani, Chem. Rev. \textbf{108}, 4072
(2008).

\bibitem{HEATH}
K. C. Beverly, J. F. Sampaio and J. R. Heath, J. Phys. Chem. B
\textbf{106}, 2131 (2002).

\bibitem{PUDDLES1}
E. Dagotto, Science \textbf{309}, 257 (2005). M. Uehara et al.,
Nature \textbf{399}, 560 (1999).


\bibitem{CB}
S. Chen et al., Science \textbf{280}, 2098 (1998). A. W. Snow and H.
Wohltjen, Chemistry of Materials \textbf{10}, 947 (1998). B. Abeles
et al., Adv. Phys. \textbf{24}, 407 (1975). P. E. Trudeau, A.
Escorcia and A.-A. Dhirani, J. Chem. Phys. \textbf{119}, 5267
(2003).

\bibitem{KONDO}
L. I. Glazman and M. E. Raikh, JETP Lett. \textbf{47}, 452 (1988).
J. Park et al., Nature \textbf{417}, 722 (2002). W. Liang et al.,
Nature \textbf{417}, 725 (2002).

\bibitem{Haldane}
F. D. M. Haldane, Phys. Rev. Lett. \textbf{40}, 416 (1978).

\bibitem{BADMETAL}
P. B. Allen, R. M. Wentzcovitch, W. W. Schulz and P. C. Canfield,
Phys. Rev. B \textbf{48}, 4359 (1993). M. Qazilbash et al., Phys.
Rev. B \textbf{74}, 205118 (2006). V. J. Emery and S. A. Kivelson,
Phys. Rev. Lett. \textbf{74}, 3253 (1995).  V. J. Emery, S. A.
Kivelson and J. M. Tranquada, Proc. Natl. Acad. Sci. \textbf{96},
8814 (1999). O. Gunnarsson and J. E. Han, Nature \textbf{405}, 1027
(2000).

\bibitem{mean free path}
O. Gunnarsson, M. Calandra, and J. E. Han, Rev. Mod. Phys.
\textbf{75}, 1085 (2003).

\bibitem{HETERO}
K. K. Gomes et al., Nature \textbf{447}, 569 (2007). L. Jinho et
al.,  Nature \textbf{442}, 546 (2006). K. M. Lang et al., Nature
\textbf{415}, 412 (2002). S. H. Pan et al., Nature \textbf{413}, 282
(2001).



\end{thebibliography}
\end{document}